\documentclass{aa}  

\usepackage{graphicx}
\usepackage{txfonts}
\newcommand{\degrees}{\ensuremath{^\circ}}
\newcommand{\hours}{\ensuremath{^\mathrm{h}}}
\newcommand{\minutes}{\ensuremath{^\mathrm{m}}}
\newcommand{\seconds}{\ensuremath{^\mathrm{s}}}

\begin{document} 

\titlerunning{NGC\,4993: constraints on the galactic merger}
   \title{NGC\,4993, the shell galaxy host of GW170817: constraints on the recent galactic merger.}

   \author{I. Ebrov\'{a}
          \inst{1}
          \and
          M. B\'{i}lek\inst{2,3}
          \and
          M.~K. Y{\i}ld{\i}z\inst{4,5}
          \and
          J. Eli\'{a}\v{s}ek\inst{6}
          }

   \institute{Nicolaus Copernicus Astronomical Center, Polish Academy of Sciences, Bartycka 18, 00-716 Warsaw, Poland\\
              \email{ebrova.ivana@gmail.com}
        \and
             Astronomical Institute,  Czech Academy of Sciences, Bo\v{c}n\'{i} II 1401/1a, 141\,00 Prague, Czech Republic
        \and
             Universit\'e de Strasbourg, CNRS, Observatoire astronomique de Strasbourg (ObAS), UMR 7550, 67000 Strasbourg, France
        \and 
             Astronomy and Space Sciences Department, Science Faculty, Erciyes University, Kayseri, 38039 Turkey
        \and
         Erciyes University, Astronomy and Space Sciences Observatory Applied and Research Center (UZAYB\.{I}MER), 38039, Kayseri, Turkey
        \and
         Institute of Theoretical Physics, Faculty of Mathematics and Physics, Charles University, 180\,00 Prague, Czech Republic
         }

   \date{Received ; accepted }

 
  \abstract
   {NGC\,4993 is the shell galaxy host of the GRB170817A short gamma-ray burst and the GW170817 gravitational-wave event produced during a binary-neutron-star coalescence.}
   {The galaxy shows signs, including the stellar shells, that it has recently accreted a smaller, late-type galaxy. 
The accreted galaxy might be the original host of the binary neutron star.}
   {We measured the positions of the stellar shells of NGC\,4993 in an HST/ACS archival image and use the shell positions to constrain the time of the galactic merger.}
   {According to the analytical model of the evolution of the shell structure in the expected gravitational potential of NGC\,4993, the galactic merger happened at least 200\,Myr ago, with a probable time roughly around 400\,Myr, and the estimates higher than 600\,Myr being improbable. 
This constitutes the lower limit on the age of the binary neutron star, because the host galaxy was probably quenched even before the galactic merger, and the merger has likely shut down the star formation in the accreted galaxy. 
We roughly estimate the probability that the binary neutron star originates in the accreted galaxy to be around 30\,\%.}
   {}


   \maketitle
%

\section{Introduction} \label{sec:intro} 

The galaxy \object{NGC\,4993} is the host of the electromagnetic counterpart \citep[SSS17a,][]{sss17a} of the gravitational-wave event \object{GW170817} \citep{gw17}. 
The gravitational waves were accompanied by a short gamma-ray burst (sGRB), \object{GRB170817A}, and followed by the ultraviolet, optical, and near-infrared emission. The event is explained as a binary-neutron-star (BNS) coalescence.

The galaxy NGC\,4993 is calssified as an early type. 
\cite{4993decam} argue that the properties they measured, qualify the galaxy as an atypical sGRB host (see their Figure\,2).
Similarly, \cite{fo17} conclude that NGC\,4993 is superlative in terms of its large luminosity, old stellar population age, and low star formation rate compared to the previous sGRB hosts.
On the other hand, \cite{my17} find NGC\,4993 to be at the old end of the range of stellar ages, but similar to some host galaxies of sGRBs. 
According to \cite{4993env}, NGC\,4993 is consistent with the distributions seen for sGRBs, although the offset of the GW170817 source position, with respect to the galactic center and normalized by the effective radius of the galaxy, is closer than about 90\,\% of sGRBs.
All three works agree that there is no, or very little, ongoing star formation in the galaxy.
However, the situation for NGC\,4993 is a bit more complicated.
The galaxy shows several signs that it underwent a relatively recent galactic merger\footnote{In this paper, the term `merger' always refers to the merger of galaxies, not to the BNS coalescence}. Thus, the BNS may have come from the secondary galaxy that had different properties. 
The following evidence suggests that the accreted secondary was a smaller late-type galaxy.

Also known as the shell galaxy MC\,1307--231 \citep{mc83}, NGC\,4993, possesses several stellar shells already visible in Digitized Sky Survey (DSS) images and even more pronounced in the recent Hubble Space Telescope (HST) archival data.
Shell galaxies account for roughly 10\,\% of early-type galaxies \citep[e.g.,][]{mc83,at13}. 
Their origin and properties are well-explained by a model where the shells are made of stars from a galaxy accreted on the host on a highly eccentric orbit \citep[e.g.,][]{q84, dc86, hq88, e12sg, bilek15, illsg17}.

The HST image also uncovers a rich dust structure around the center of the galaxy.
Its irregular appearance suggests that the dust is not in hydrostatic equilibrium, indicating a recent merger.
Also, the possible low-luminosity active galactic nucleus \citep{4993decam,4993env,4993jet} could have been triggered by the gas brought to the central region by the intruding late-type galaxy. 
Further evidence that the accreted galaxy was a gaseous late-type galaxy was reported in \cite{4993decam}. The systematic change of the S{\'e}rsic index and position angle of the fitted light profile for different bands suggests that there could be two stellar populations overlaid with different orientations. Moreover, their spectral energy distribution fits prefer younger stellar ages in the outer regions. These regions seem to be associated with the shells \citep[see Figure\,4 of][]{4993decam}.

The majority of the stellar population of NGC 4993 is old \citep{4993decam}. The younger population has a different spatial distribution and orientation compared to the older stars. The younger stars are also associated with the shells. All these aspects indicate that the accreted galaxy brought the young stars, and that the star formation in the primary galaxy was already quenched some time before the galactic merger.
Since the time distribution of the coalescence events of binary neutron stars is expected to decline steeply with the stellar age \citep[e.g.,][]{bns1,bns2,bns3}, there is an increased probability that progenitor of GW170817 originated in the accreted galaxy (see Sect.\,\ref{ssec:dis-pbt}). Therefore, the environment, the type of the galaxy, and the galactocentric distance at which the progenitor was born, could be very different from what is derived from the current observations. 

In the galactic-merger scenario, shells are formed by the stars currently located near the apocenters of their very eccentric orbits. 
The shells emerge after the first close pericentric passage of the galaxies and can easily survive several gigayears \citep[e.g.,][]{e12sg,illsg17,ma19} and possibly even longer than a Hubble time \citep{dc86}.
The unique shell kinematics makes it possible to connect the gravitational potential of the galaxy, the age of the shell system, and galactocentric positions of the shell edges \citep{q84, dc86, e12sg, bilek13, bilek14}. 
We use the shell-system age (i.e., the time that elapsed since the stars that form the shell system were stripped from the accreted galaxy) as a proxy for the merger time. 
In the event that the secondary galaxy survives more pericentric passages, and therefore leads to more shell systems with different ages in the host galaxy (see Sect.\,\ref{ssec:dis-2g}), we adopt the age of the youngest shell system.

Here, we use publicly available images and data from the literature to measure shell positions, construct the gravitational potential, and infer the shell-system age. It would be hard to measure the star formation history of the secondary galaxy since the galaxy is now dissolved and overlaid on the body of a more massive galaxy. 
The shell-system age can serve as an estimation of the time when the star formation in the secondary galaxy probably stopped as a result of the disintegration of the accreted galaxy.
Our estimate can provide constraints on the models of the evolution of the binary neutron star.

\begin{figure*}
\resizebox{\hsize}{!}{\includegraphics{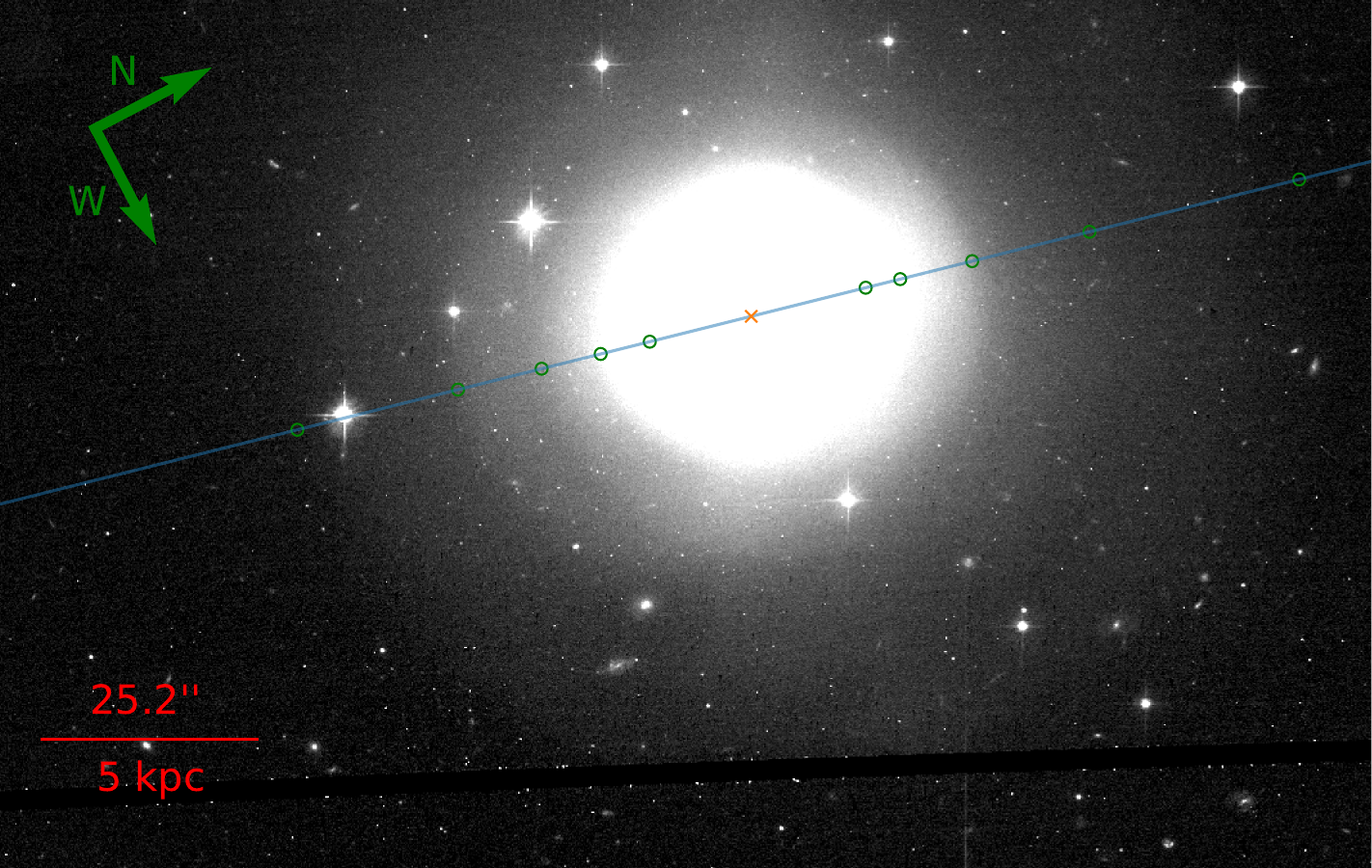}\includegraphics[width=1.055\hsize]{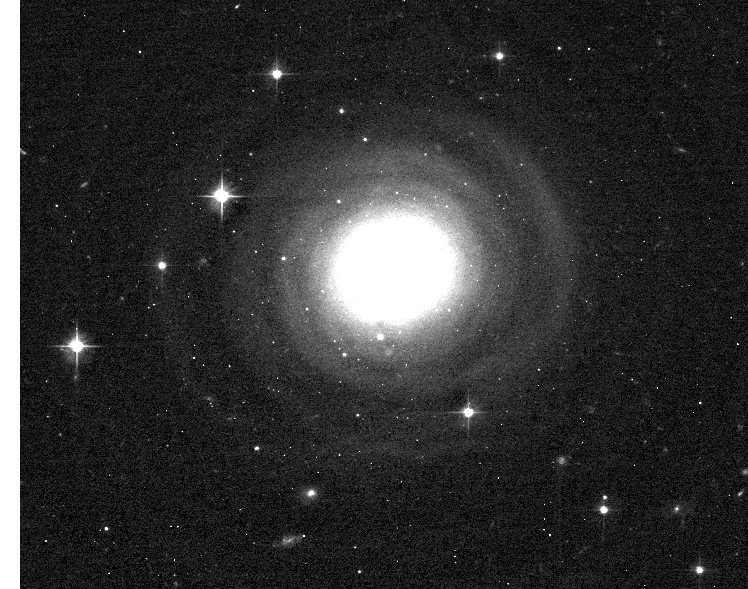}} 
    \caption{
Image of shell galaxy NGC\,4993 (based on the HST/ACS data in the F814W filter).
\textit{Left}: The image processed by intensity scaling to show the outer shells. 
The blue line indicates the visually determined major photometric axis. 
The orange cross denotes the center of the galaxy. 
The green circles on the blue line mark the positions of the measurements of shell radii (Table\,\ref{tab:shells}).
\textit{Right}: The image processed by minimum masking technique \citep{bilek16}, with the filter size of 1.2\arcsec,~reveals the shells close to the galaxy center. We note the two-arm spiral-like morphology of the shells.
}
    \label{fig:4993}
\end{figure*}

\section{Shells of NGC 4993} \label{sec:sh} 

We adopt the NGC\,4993 distance of $41.0\pm3.1$\,Mpc -- the combined redshift and fundamental-plane value from \cite{4993dis}. 
The value is consistent with the surface-brightness-fluctuation distance to NGC\,4993 derived in \cite{4993dis2}.
One arcsecond corresponds to $0.20\pm0.02$\,kpc.

We used the publicly available Advanced Camera for Surveys (ACS) F606W Hubble Space Telescope (HST) image of NGC\,4993 obtained on Apr\,28, 2017 (exposure time 696\,s).
The outermost detected shell is near the edge of the HST image.
We also examined the publicly available data from Dark Energy Camera (DECam) taken on Jun 04 -- Jun\,09, 2015 (360\,s in \textit{r} and 100\,s in \textit{g} filters). The DECam is installed at a four-meter telescope at the Cerro Tololo Inter-American Observatory (CTIO) in Chile, and it has a field of view of roughly $2.2\degrees$ in diameter. We found no additional shells in the DECam image; thus, all following measurements of the shells are performed on the HST data.

We detected the shells visually in the HST/ACS data. We present the HST image of the galaxy in Fig.\,\ref{fig:4993}.
We can detect all shells either via a linear scaling of the image or after applying the minimum masking technique \citep[see Appendix\,B in][]{bilek16} with the filter size of 1.2\arcsec (right-hand panel of Fig.\,\ref{fig:4993}).
The minimum masking has improved the visibility of some of the shells, particularly those closer to the galaxy center.
The inner shell structure appears to have a two-arm spiral-like morphology. 
Such a morphology of inner shells is observed in many other Type\,II shell galaxies \citep{wil87,pri90}, for instance, in MC\,0422--476 \citep{wil00}.
The minimum masking obscures the sharp-edged nature of the inner shells. The sharp edges of the inner shells of NGC\,4993 can be seen, for example, in Figure\,1 of \cite{my17}, and they distinguish them from the classical spiral arms. 

To constrain the shell-system age, we need to measure the shell radii. 
This is ambiguous given the spiral-like morphology of the shells. 
Inspired by the axially symmetric shell systems (see Sect.\,\ref{sec:evo}), we measured the shell radii at the points where the shell edges cross the visually determined major axis of the galaxy, as indicated in the left-hand panel of Fig.\,\ref{fig:4993}. 
A further uncertainty of the measured shell radii comes from the blurriness of the shell edges. We estimate this uncertainty to be smaller than $\pm5$\,\% of the shell radius for all the shells.
We list the adopted shell radii in Table\,\ref{tab:shells}. The shells lying north (south) of the galaxy center have positive (negative) signs.
The center was set to the coordinates of the brightest pixel in the central region of the galaxy (RA = $13\hours09\minutes47.7\seconds$, Dec = $-23\degrees23\arcmin02.2\arcsec$).
The electromagnetic counterpart of GW170817 was observed $10.3\arcmin$ from that center, meaning at slightly lower radius than the innermost observed shell. 
Given the approximative character of the calculation of shell-system age (Sects.\,\ref{sec:evo} and \ref{ssec:dis-type2}), the uncertainty in determining the photometric axis and the center of the galaxy is probably unimportant.

\begin{table}
    \centering
    \caption{Measured shell-edge radii of NGC\,4993}
    \begin{tabular}{ccc} 
        \hline
Shell radius & Detection method &\\
        \hline
$-12$\arcsec & minimum masking & \\
$~14$\arcsec & minimum masking &  \\
$~18$\arcsec & linear scaling & \\
$-18$\arcsec & linear scaling & \\
$-25$\arcsec & linear scaling & \\
$~27$\arcsec & linear scaling & \\
$-35$\arcsec & linear scaling & \\
$~41$\arcsec & linear scaling &  \\
$-55$\arcsec & linear scaling & \\
$~66$\arcsec & linear scaling & \\
        \hline
    \end{tabular}
    \label{tab:shells}
\end{table}

\section{Model for the gravitational potential} \label{sec:gp} 

The evolution of the shell structure depends on the gravitational potential of the galaxy. We constructed the model of the gravitational potential of NGC\,4993 using the measurements of the stellar component from literature and coupling them with the dark matter component via the abundance matching technique. 

We modeled the stellar component as a deprojected S{\'e}rsic profile assuming spherical symmetry. The parameters of the S{\'e}rsic profile were taken from \cite{4993decam}. 
We used their GALFIT output for the \textit{r}-band DECam image \citep[see Table\,1 in][]{4993decam}: the apparent magnitude of 11.90, the effective radius of 16.74\arcsec, and the S{\'e}rsic index of 3.7.
Similar values of the fit were also derived in \cite{my17}.
Furthermore, we assumed a constant \textit{r}-band mass-to-light ratio of 5.23 derived by \citet{4993decam} from a 6dF spectrum of NGC\,4993.

We estimated the mass of the dark matter halo by using Equation\,(15) from \cite{am06} -- the relation between the central stellar velocity dispersion and the halo mass derived by the abundance matching technique. The value of the central velocity dispersion, $160.0\pm9.1$\,km\,s$^{-1}$, was inferred from the 6dF spectrum by \cite{4993decam}. This value is roughly consistent with the weighted mean value, $\langle \mathrm{log}(\sigma_{0}/\mathrm{km\,s^{-1}}) \rangle =2.241\pm0.012$, compiled from measurements in literature by \cite{my17}.
The central velocity dispersion of $160.0$\,km\,s$^{-1}$ corresponds to the halo mass of $M_{\rm halo}=194\times10^{10}$\,M$_{\sun}$.
The uncertainties in the velocity-dispersion measurements translate to the errors in the inferred halo mass of $+35$ and $-31\times10^{10}$\,M$_{\sun}$.
Uncertainties of Equation~(15) from \cite{am06} are indicated in their Figure\,3c. For the given value of the NGC\,4993 velocity dispersion, the errors of the halo mass are roughly $+120$ and $-70\times10^{10}$\,M$_{\sun}$.
We modeled the dark matter halo using the Navarro--Frenk--White (NFW) profile \citep{nfw97}. 
The mass of the NFW halo constrains its concentration.
The relation derived in \cite{dut14} leads to the concentration parameter $c_{200}=7.52$. 
The intrinsic scatter of this concentration-mass relation 0.11\,dex corresponds to the $c_{200}$ errors of $+2.17$ and $-2.07$.

\begin{figure*}
\resizebox{0.98\hsize}{!}{\includegraphics{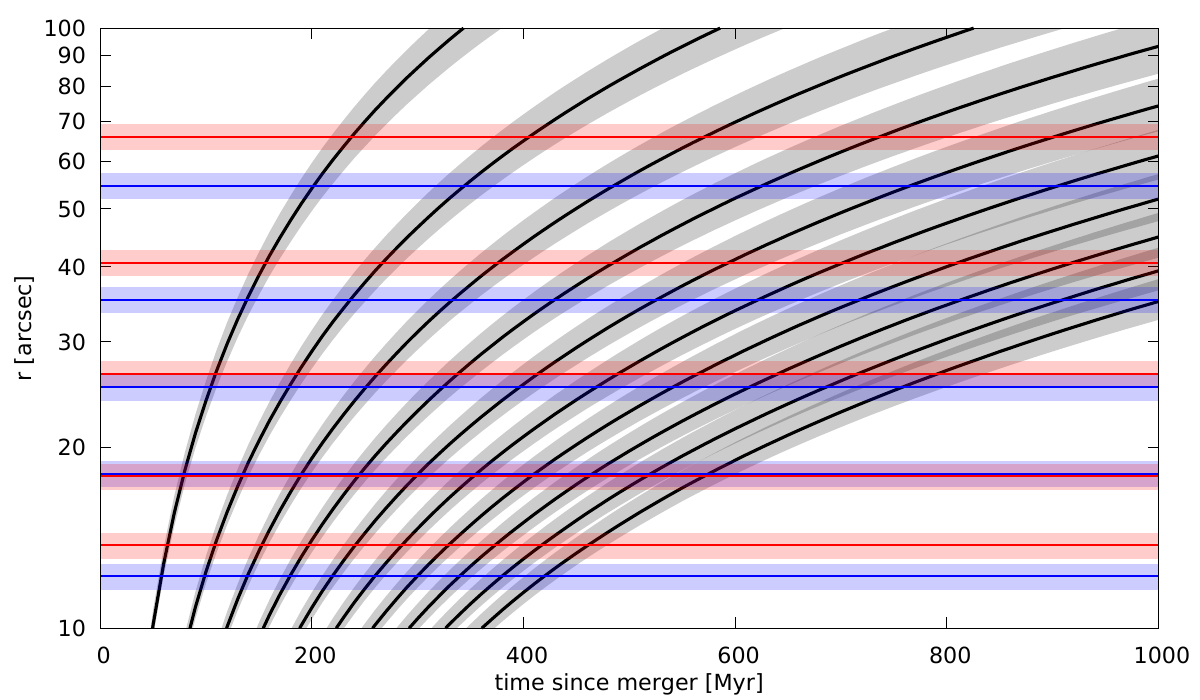}} 
    \caption{Black curves show the modeled evolution of the radii for the first ten shells in the gravitational potential of NGC\,4993. The gray transparent regions correspond to the uncertainties in the halo mass and concentration.
Red (blue) lines correspond to the actual shell edges north (south) of the center of NGC\,4993. The transparent red and blue stripes correspond to the measurement errors of the shell radii, $\pm5$\,\%, measured on the major axis of the galaxy.}
    \label{fig:evo}
\end{figure*}

\section{Evolution of shell radii} \label{sec:evo}

Methods for the analytical computation of the evolution of shell radii, in a given gravitational potential, were developed only for Type\,I shell systems, see \cite{q84, dc86, e12sg, bilek13, bilek14}.
Type\,I shell systems are symmetric around the major photometric axis of the galaxy, and the shells are almost circular arcs centered on the galaxy core \citep{wil87,pri90}.
This model assumes that the secondary galaxy was accreted on the host on a radial trajectory and that the stars released from the accreted galaxy move also on strictly radial orbits. 
Shells emerge on the radii, where the stars are currently located near the apocenters of their orbits -- so-called current turning points, $r_{\mathrm{TP}}(n)$, where $n$ is a serial number of a given shell.
The evolution of $r_{\mathrm{TP}}$ is given by Equations~(1) and (2) of \cite{e12sg}, originally published in \cite{q84}:
\begin{equation}
t=(n+1/2)T(r_{\mathrm{TP}}),
\label{eq:Trn}
\end{equation}
where $t$ is the time elapsed since stars were released near the center of the host galaxy. $T(r)$ is the period of radial motion at a galactocentric radius $r$ in the host galaxy potential $\phi(r)$:
\begin{equation}
T(r)=\sqrt{2}\int_{0}^{r}\left[\phi(r)-\phi(r')\right]^{-1/2}\mathrm{d}r'.
\label{eq:Tr}
\end{equation}
In our case, the host galaxy potential $\phi(r)$ is given by the distribution of the dark and baryonic matter derived in Sect.\,\ref{sec:gp}. 
At the given time $t$, the $n$th current turning point occurs at $r_{\mathrm{TP}}$ that satisfies Eq.~(\ref{eq:Trn}) for the given $n$. 
We applied a small additional correction to $r_{\mathrm{TP}}$, reflecting that the actual shell radius is located on a slightly higher radius due to the non-zero phase velocity of the shells.
This correction is based on Equations~(12) and (13) of \cite{e12sg} approximating the shell phase velocity by the velocity of the current turning point, Equations~(3) of \cite{e12sg}, which is very similar \citep[see Table~1 of][]{e12sg}.
For more details, see also Sections\,9.1 and 9.2 in \cite{phd} or Section\,2.4.1 in \cite{phdmb}. 
See also \cite{bilek13, bilek14}, where those methods were applied to the shell system of the galaxy NGC\,3923.
We note that the assumption on the radial orbits is broken to some degree for the nonaxisymmetric shell system of NGC\,4993. In Sect.\,\ref{ssec:dis-type2}, we argue that the model still gives a useful estimate of the merger time for the galaxy.

The black curves in Fig.\,\ref{fig:evo} show the evolution of the radii for the first ten shells in the gravitational potential of NGC\,4993. 
Since the outermost modeled shell corresponds to the shell that was created first, shells are traditionally numbered from outside to inside.\footnote{\label{fn:sh0}Here, we do not consider the so-called zeroth shell. This shell occurs around apocenters of the stars that were just released from the secondary galaxy and performed less than a half of one orbit in the host galaxy. Such a shell is very diffuse, and it is a part of the cloud of tidal debris loosened from the secondary, because these stars are not yet folded in the phase plane. Some authors do not even consider it to be a true shell \citep{hq88}. Moreover, the zeroth shell moves quickly and disappears from the system very fast. The appearance of the outermost observed shell in NGC\,4993 is not consistent with the zeroth shell.}
The leftmost curve corresponds to the shell number one, which consists of the stars reaching apocenters of their first full orbital period in the host galaxy.
The shell radius increases with time, and more shells emerge. 
Each subsequent shell is formed by stars that have completed more orbital periods, and thus, it expands slower than the previous shell at the same radius. 
The older the shell system is, the more shells it possesses. 

The gray regions in Fig.\,\ref{fig:evo} indicate the combined uncertainties of the modeled shell positions. The shell-position uncertainties come from the uncertainties in the halo mass and concentration given by the measurement error of the central velocity dispersion and by the intrinsic scatter of the stellar-to-halo mass relation and the concentration-mass relation for the NFW halo, see Sect.\,\ref{sec:gp}. 
Note that the uncertainties in the evolution of shell radii are not independent: if the halo mass and concentration is different from the values used in calculations of the thick black lines in Fig.\,\ref{fig:evo}, all the curves would be shifted in the same direction.
The real uncertainties of the model are probably bigger since the model was not constructed to account for the shell structure with the morphology observed in NGC\,4993 (Type\,II shells); see Sect.\,\ref{ssec:dis-type2}.

The shell radii measured in the HST image of NGC\,4993 (Sect.\,\ref{sec:sh}, Table\,\ref{tab:shells}) are indicated by the horizontal lines in Fig.\,\ref{fig:evo}. Red (blue) lines correspond to the shells north (south) of the center of the galaxy.

\section{Discussion} \label{sec:dis}

Here, we infer the probable time since the galactic merger and discuss the limitations and relevance of our estimate. 

\subsection{Type\,II shell system} \label{ssec:dis-type2}

The method described in Sect.\,\ref{sec:evo} was, in a more complex form, applied to a rich Type\,I shell system of the galaxy NGC\,3923 in \cite{bilek13, bilek14}.
\cite{wil87} and \cite{pri90} recognized three different morphological categories of shell galaxies.
Type\,I shells are well-aligned with the major axis of the galaxy in a biconic structure, interleaved in the radius (i.e., the next outermost shell is usually on the opposite side of the galactic center), with the shell separation increasing with the galactocentric radius. Such shells are well-reproduced in models involving a radial galactic merger.
Type\,II shells appear randomly distributed all around a rather circular galaxy. 
Shell systems that have more complex structures, or have too few shells to be classified, are put into a Type\,III category. 
Shell galaxies of all three categories are observed in roughly equal numbers \citep{pri90}. For more details, see also Sections\,3.3 in \cite{phd}.

The observed Type\,II morphology of the shells in NGC\,4993 indicates that the galaxy accreted a smaller disky galaxy on a highly eccentric but nonradial trajectory, with the disk probably inclined with respect to the orbital plane of the galactic merger, see, for example, Figure\,3 of \cite{q84}, or Figure\,11 of \cite{dc86}.
Such shells are not well-reproduced by the idealized model of a radial galactic merger with stars on radial orbits, but there is no equivalent model for the Type\,II systems.
It would require more detailed $N$-body simulations to account for such a merger and to infer the merger time with higher accuracy. There are two major obstacles preventing a precise reproduction of the observed distribution of shell radii by the model in use:

(1) The positions of the true shell edges probably lag behind the modeled radii, since the stars on nonradial orbits have longer orbital periods for the same apocentric distances.
The period of an oscillation in the radial direction is the longest for the stars on nearly circular orbits. 
Using the epicyclic approximation \citep{bt}, we derived that, in the potential of NGC\,4993, the period for such particles is 1.5--1.6 times longer than for the particles on radial orbits, in the range of apocentric distances corresponding to the range of the observed shells. 
This means that the assumption of the radial orbits leads to the underestimation of the shell age by at most 50--60\,\%.
The actual error caused by the orbit nonradiality is probably much smaller, since simulations show that the formation of shells of any type requires a highly eccentric galactic collision \citep[e.g.,][]{illsg17}.
We performed a collisionless $N$-body simulation of a galactic merger resulting in a Type\,II shell system, see Appx.\,\ref{apx:sim}. 
We compared the evolution of the shell radii in the simulation with the analytical model calculated for the simulated galaxy (Appx.\,\ref{sapx:evo}).
Generally, the model agrees well with the simulation (see Fig.\,\ref{fig:hist}), especially for the outer shells, where the difference is only a few percent of the shell radius.
The projection effects seem to be the biggest source of the uncertainties for the simulated galaxy (Appx.\,\ref{sapx:proj}) -- the shell radius and visibility depend on the viewing angle. The maximal alteration of the radius is between 2\,\% and 19\,\% for the shells in the simulated galaxy.

(2) Type\,II shells do not lie at one axis, but each one has its midpoint at a different azimuth, which cannot be determined without more detailed data processing, and a more advanced model. 
Since, in NGC\,4993, the radius of a shell depends on the azimuth, measuring the shell radius on the symmetry axis introduces a certain systematic error. The true shell midpoint, in most cases, probably lies somewhere between the radii of two subsequent shells measured on the major axes of the galaxy. This is probably the reason why the measured shell spacing does not increase monotonically with the radius, in contradistinction to the model prediction. 
Furthermore, since the true midpoints of two subsequent shells change their azimuths by an angle different from $180\degrees$, we cannot expect the shells of NGC\,4993 to be regularly interleaved in radius. Thus we do not assign sides to the modeled shells \citep[as in][]{bilek13, bilek14, bilek15}, even though the model expects them to be interleaved. Moreover, it is expected that one or more of the outermost shells of NGC\,4993 may be too faint to be observed in the current images or completely vanished due to the lack of stars with sufficiently high orbital energy. If none or an even number of the outermost shells are not observed, then modeled shells with odd (even) numbers should be assigned to the north (south) side of the galaxy, and vice versa.

For the reasons mentioned above, in the case of NGC\,4993, our model of the shell system evolution is not supposed to exactly reproduce the observed distribution of shell radii at any particular time.
Nevertheless, we still can put significant constraints on time since the merger of the galaxies, using the following knowledge:  
(i) The modeled evolution of the outermost shell basically corresponds the shortest possible time in which stars, released near the galactic center, could have reached the given position in the given gravitational potential of the host galaxy. 
This sets a lower limit for the merger time of the galaxies.
(ii) For a Type\,II shell system, the rate at which shells are generated is expected to be slower, but well-approximated by the model (see the point (1) above). This means that at the time corresponding to the shell-system age, the number of observed and modeled shells, in a certain range of radii, should be comparable, even though the exact positions of the shells are not reproduced.

\subsection{Single-generation scenario} \label{ssec:dis-1g}

Firstly, we assume that all ten observed shells are made of stars that have been stripped from the secondary galaxy during the same pericentric passage, meaning the shell system of NGC\,4993 consists of one shell generation. 
This assumption is consistent with simulations, where up to ten shells are produced in the cases where the secondary galaxy is forced to dissolve during the first pericentric passage \citep{dc86,e12sg}.
In the given model, the outermost detected shell of NGC\,4993 cannot reach its measured radius of 66$^{\prime\prime}$ (13.0\,kpc) sooner than 200\,Myr after the stars were stripped.
Thus, 200\,Myr can be set as a safe lower estimate of the merger time. 

At around 400\,Myr after the merger, the radius of the outermost observed shell corresponds to the position of the second shell in the model. 
Moreover, at this time, the number of the observed and predicted shells is comparable in the region occupied by the observed shells. 
This would mean that the outermost observed shell is actually the shell number two. 
In such a situation, deeper photometric images of NGC\,4993 could reveal the unobserved outer shell. 
Such an observation would help to improve the accuracy of the inferred time of the merger of the galaxies. 
In the event that the outer shell is not found, it may just mean that the first shell already vanished due to the lack of stars with sufficiently high orbital energy.
For times around 600\,Myr and higher, the model predicts too many shells that are too densely placed (especially for the inner parts) to account for the observed shells of NGC\,4993. 
It should be noted that only the evolution of the first ten shells is displayed in Fig.\,\ref{fig:evo}.
It is unlikely that some shells escape observations in the region occupied by the six innermost shells, since they form a continuous spiral structure.
In this single-generation scenario, the galactic merger thus occurred at most 600\,Myr ago.

\subsection{Two-generation scenario} \label{ssec:dis-2g}

Now, we consider a two-generation shell structure for NGC\,4993. That means that the first shell generation was released during the first pericentric passage, but the secondary galaxy preserved a significant amount of stellar matter, and it was able to produce a second shell generation during the second pericentric passage \citep{hw90,katka11}.
A multi-generation scenario seems to be necessary to explain the rich Type\,I systems like NGC\,3923 \citep{dc87,bilek13}, which has the observed ratio between the outermost and innermost shell radii at least 65 \citep{bilek16}. This ratio is only about 5.4 for NGC\,4993. Simulations are able to reproduce the ratio up to the value of around six in a single shell generation \citep{dc87}, but the two-generation scenario is still possible.
In such a scenario, the first shell generation occurs at higher radii than the subsequent generation due to the loss of the orbital energy of the surviving secondary galaxy via dynamical friction.

Given the two-arm spiral-like morphology that binds the six inner shells of NGC\,4993 together \citep[as observed for the inner shells of many other Type\,II shell galaxies;][]{wil87,pri90}, it is unlikely that these shells come from different generations. For the same reason, it is unlikely that an undetected shell, which would have to violate the spiral-like morphology, is present in the inner region. If the six inner shells constitute the second generation, the age of this generation would still be around 400\,Myr, since it is impossible to produce six shells within the radial range of the six observed inner shells earlier than this time.
If the four outermost observed shells are assigned to the first generation, its age would be around 600\,Myr, or, alternatively, up to around 800\,Myr if there is an undetected shell between the observed ones. 
In this scenario, 2-4 outermost shells of the first generation are not observed, and there are not enough stars in the first shell generation with sufficiently low orbital energy to produce visible shells at radii around $30$\arcsec and lower. For the second generation, four outermost shells are not observed.

The GW170817 electromagnetic counterpart was observed slightly under the radius of the innermost observed shell. It is then more probable that the binary-neutron-star (BNS) progenitor is connected to the second shell generation if the BNS originated in the secondary. Moreover, if the secondary galaxy, after the first pericentric passage, maintained enough matter to produce the second shell generation, it could still have been forming stars in the meantime. Thus, even in the case of a two-generation scenario, the galactic merger again occurred at around 400\,Myr, and if the BNS progenitor had originated in the secondary galaxy, it would probably have been released at this time.

\subsection{Shell-system age} \label{ssec:dis-age}

According to the model, in both cases (the single- and two-generation scenario), the final stages of the galactic merger probably occurred around 400\,Myr.
Our lower estimate of the merger time 200\,Myr is in the contradiction with \cite{4993decam} claiming the merger time to be lower than 200\,Myr. The estimate calculated in Section~5.1 in \cite{4993decam} is based on an incorrect assumption that the time since the merger is shorter than the crossing time. In fact, the nature of the shells implies that the crossing time at the radius of the outermost shell is rather close to the time since the merger or lower if the first shell or shells (that were created after the merger) are not observed. 

\subsection{The original galaxy host of the BNS} \label{ssec:dis-pbt}

Here, we derive a broad estimate of the probability that the BNS progenitor of the gravitational-wave event originates in the accreted secondary galaxy. 
The delay-time distribution of the BNS coalescence is expected to follow $t^{-1}$, where $t$ is the time since the BNS formation, and the minimum delay time between the binary formation and coalescence is assumed to be around 20\,Myr \citep{bns1,bns2,bns3}. The minimum delay time is smaller than the expected accuracy of our estimate of the galactic-merger time, and we can neglect it.
Hence, the probability, $P_{1}$, that the BNS come from a population of stars with mass $M_{1}$ and age $t_{1}$ is simply 
\begin{equation}
P_{1}=\frac{M_{1}/M_{\mathrm{tot}}}{t_{1}}\frac{1}{C},
\label{eq:p}
\end{equation}
where $M_{\mathrm{tot}}$ is the total mass of all stellar populations in the galaxy, and $C$ is a normalization, ensuring that the overall probability is equal to one.

To estimate the probability, we need to assume some stellar-mass ratios of the galactic merger that created the shells in NGC\,4993. 
\cite{illsg17} study shell galaxies in the cosmological simulation Illustris and find that they form preferentially through mergers with the stellar-mass ratio 1:10 or higher.
\cite{ssp}, in their sample of 214 shell galaxies from the Hyper Suprime-Cam Subaru Strategic Program, examine the color differences between the shells and their host galaxies and conclude that mergers with the mass ratios around 1:4 or lower dominate the majority of their sample. They also argue that the major mergers more likely lead to the formation of the Type\,I shell galaxies. 
We used the HST image of NGC\,4993 to estimate the ratio. We modeled the light of the main body of the host galaxy and subtracted it from the image. We compared the signal in the residual image with the values recovered by the same procedure for the simulated galaxy. 
The simulation is not meant to precisely reproduce the shell system of NGC\,4993 but the gravitational potential of the primary galaxy at the beginning of the simulation corresponds to the potential infrared for NGC\,4993.  
The residual image of the simulated galaxy contains about a half of the light coming from the particles of the secondary galaxy in the measured area. 
This leads to the stellar-mass ratio of 1:3 for the merger that occurred in NGC\,4993. Details of the procedure are described in Appx.\,\ref{apx:fits}.

Lastly, we need to assume an age distribution of the stellar populations in the progenitors of the galactic merger.
We took advantage of the analysis of 436 galaxies of the CALIFA survey published in \cite{califa17}. Figure\,10 of \cite{califa17} shows the average stellar-mass fraction in seven age bins (11.4--0.7\,Gyr) for galaxies of different Hubble types and stellar masses. 
We used the CALIFA data points measured for the whole galaxy up to 2 half-light radii. 
The primary (secondary) galaxy was represented by an E-type (Sbc-type) galaxy in the mass bin of $10^{11}$--$10^{11.3}$\,M$_{\sun}$ ($10^{9.8}$--$10^{10.6}$\,M$_{\sun}$).

For the chosen merger mass ratio and the age distribution of the stars, the probability that the BNS progenitor of the gravitational-wave event originates in the accreted galaxy is 33\,\%. 
Since the time distribution of the BNS coalescence declines steeply with time, the probability would be even higher if there was a merger-related enhanced star formation in the secondary galaxy. 

\section{Summary and conclusions}

NGC\,4993 is the shell galaxy host of the GRB170817A short gamma-ray burst and the 
GW170817 gravitational-wave event produced during a binary-neutron-star coalescence. 
Stellar shells are known to originate in galactic merges.
A scenario, in which the host galaxy recently accreted a smaller late-type galaxy, may account for the existence of the shells together with the distribution of the younger stellar population, the dust lanes in the central region, and the low-luminosity active galactic nucleus.  
That implies that NGC\,4993 may not be the original host of the binary neutron star. The binary star could have been born in a very different environment in a galaxy with properties distinct from the properties measured for NGC\,4993. Specifically, the accreted galaxy had a lower total (stellar) mass and probably a higher gas fraction with an ongoing star formation, which could even have been enhanced by the tidal interaction prior to the galaxy disintegration.
Since the host galaxy was probably quenched well before the merger (see Sect.\,\ref{sec:intro}), the time of the galactic merger gives an approximate lower limit of the age of the binary neutron star at the time of its coalescence.

Here, we used the HST/ACS image to measure the galactocentric radii of the stellar shells in the galaxy. 
The DECam data show additional shells, neither inside nor outside the field of view of the HST image.
The shell system displays a spiral-like morphology, especially in the inner parts. 
The abundance matching technique and data from literature were used to construct the model of the gravitational potential of NGC\,4993.
We computed the evolution of the shell radii in this potential employing an analytical model of shells that consist of stars on radial orbits. 
We estimated the time of the galactic merger by comparing the computed and observed shell radii.

According to the model, the galactic merger happened at least 200\,Myr ago, with the most probable time roughly around 400\,Myr. Estimates higher than 600\,Myr are improbable. 

The biggest source of the inaccuracy of the estimate is probably the fact that the used model was developed for axisymmetric shell systems, which is not the case of NGC\,4993.
A better model of the gravitational potential of the galaxy and/or more detailed $N$-body simulations are needed for better estimates of the time of the galactic merger.

We estimated the probability that the binary-neutron-star progenitor of the gravitational-wave event originates in the accreted galaxy to be around 30\,\%.

\begin{acknowledgements}

We thank David A. Coulter, Ryan J. Foley, Marcin Semczuk, and Yohei Miki for valuable comments and advice.

We acknowledge support from the Polish National Science Centre under grant 2013/10/A/ST9/00023 and 2017/26/D/ST9/00449 (IE).

Based on observations made with the NASA/ESA Hubble Space Telescope, obtained from the data archive at the Space Telescope Science Institute. STScI is operated by the Association of Universities for Research in Astronomy, Inc. under NASA contract NAS 5-26555.

This project used public archival data from the Dark Energy Survey (DES). Funding for the DES Projects has been provided by the DOE and NSF (USA), MISE (Spain), STFC (UK), HEFCE (UK). NCSA (UIUC), KICP (U. Chicago), CCAPP (Ohio State), MIFPA (Texas A\&M), CNPQ, FAPERJ, FINEP (Brazil), MINECO (Spain), DFG (Germany) and the collaborating institutions in the Dark Energy Survey, which are Argonne Lab, UC Santa Cruz, University of Cambridge, CIEMAT-Madrid, University of Chicago, University College London, DES-Brazil Consortium, University of Edinburgh, ETH Z{\"u}rich, Fermilab, University of Illinois, ICE (IEEC-CSIC), IFAE Barcelona, Lawrence Berkeley Lab, LMU M{\"u}nchen and the associated Excellence Cluster Universe, University of Michigan, NOAO, University of Nottingham, Ohio State University, University of Pennsylvania, University of Portsmouth, SLAC National Lab, Stanford University, University of Sussex, and Texas A\&M University. 

Based in part on observations at Cerro Tololo Inter-American Observatory, National Optical Astronomy Observatory (NOAO Prop. ID: 2015A-0616, PI: H. Jerjen), which is operated by the Association of Universities for Research in Astronomy (AURA) under a cooperative agreement with the National Science Foundation.

\end{acknowledgements}


\bibliographystyle{aa}
\bibliography{4993} 

\appendix
\section{Simulation} \label{apx:sim}

In order to better constrain several unknown quantities in our calculations, we use a collisionless $N$-body simulation. The purpose of the simulation is not to precisely reproduce the shell system of NGC\,4993 -- that would require the exploration of a vast parameter space, which is beyond the scope of this work.

\subsection{Details of the simulation} \label{sapx:setup}

We performed a set of 15 simulations of a merger of an early-type primary galaxy with a smaller disky secondary. The simulations differ in the total mass of the secondary, in the mass and concentration of the disk, in the inclination of the disk with respect to the orbital plane or axis, and in the relative velocity and angular momentum of the merging galaxies. 
We selected a simulation that, at some stage of its evolution, most closely resembles shells of NGC\,4993 in terms of being a Type\,II shell system and having a similar number of shells in a roughly similar range of galactocentric radii. 

The $N$-body models of the merger progenitors were generated using MAGI \footnote{\url{https://bitbucket.org/ymiki/magi}} \citep{magi}.
The primary was a spherically symmetric early-type galaxy with S{\'e}rsic stellar component (with the mass of $12.24\times10^{10}$\,M$_{\sun}$, effective radius of 3.35\,kpc, and S{\'e}rsic index of 3.7) and NFW dark matter halo (with the mass of $193.9\times10^{10}$\,M$_{\sun}$ and concentration of 7.52) as derived for NGC\,4993 in Sect.\,\ref{sec:gp}.
The secondary galaxy had a disk with the mass of $1.2\times10^{10}$\,M$_{\sun}$ and the radial and vertical scale parameters of 3 and 0.6\,kpc, respectively. 
The scale parameters were chosen to be within the range of typical values for disks of the corresponding mass \citep{scale1,scale2}. 
The dark halo of the secondary was represented by an NFW sphere with the mass of $30.6\times10^{10}$\,M$_{\sun}$ and the concentration of 9.1.
The parameters of the NFW halo were derived in the same manner as in the case of the model of the gravitational potential of NGC\,4993, see Sect.\,\ref{sec:gp}.
Each component of each progenitor was generated with $10^{6}$ particles leading to the total number of $4\times10^{6}$ particles in the merger simulation.

We used the simulation code Gadget-2 \citep{sp05} with the adopted softening
scale lengths of 0.6\,kpc (0.1\,kpc) and 1.2\,kpc (0.7\,kpc) for the stellar and dark-matter component of the primary (secondary) galaxy, respectively.
Firstly, the primary (secondary) galaxy was evolved in isolation for 1\,Gyr (1.25\,Gyr). 
The disk of the secondary was inclined by $45\deg$ with respect to the collisional plane, $xy$, around both axes using the rotation matrix 
$$
\quad
\begin{bmatrix}
\sqrt{2} & 0 & \sqrt{2} \\
-0.5 & \sqrt{2} & 0.5 \\
-0.5 & -\sqrt{2} & 0.5
\end{bmatrix}
\quad,
$$
then, the secondary was placed on the $x$-axis 70\,kpc from the center of the primary with the relative velocity of 29.3\,km\,s$^{-1}$ aiming in the tangential direction.
Galaxies are fully merged shortly after four pericentric passages that occur 0.4, 0.7, 0.9, and 1.1\,Gyr after the beginning of the simulation. 
The pericentric distance of the first passage is around 6\,kpc.

\begin{figure}[ht!]
\resizebox{1.0\hsize}{!}{\includegraphics{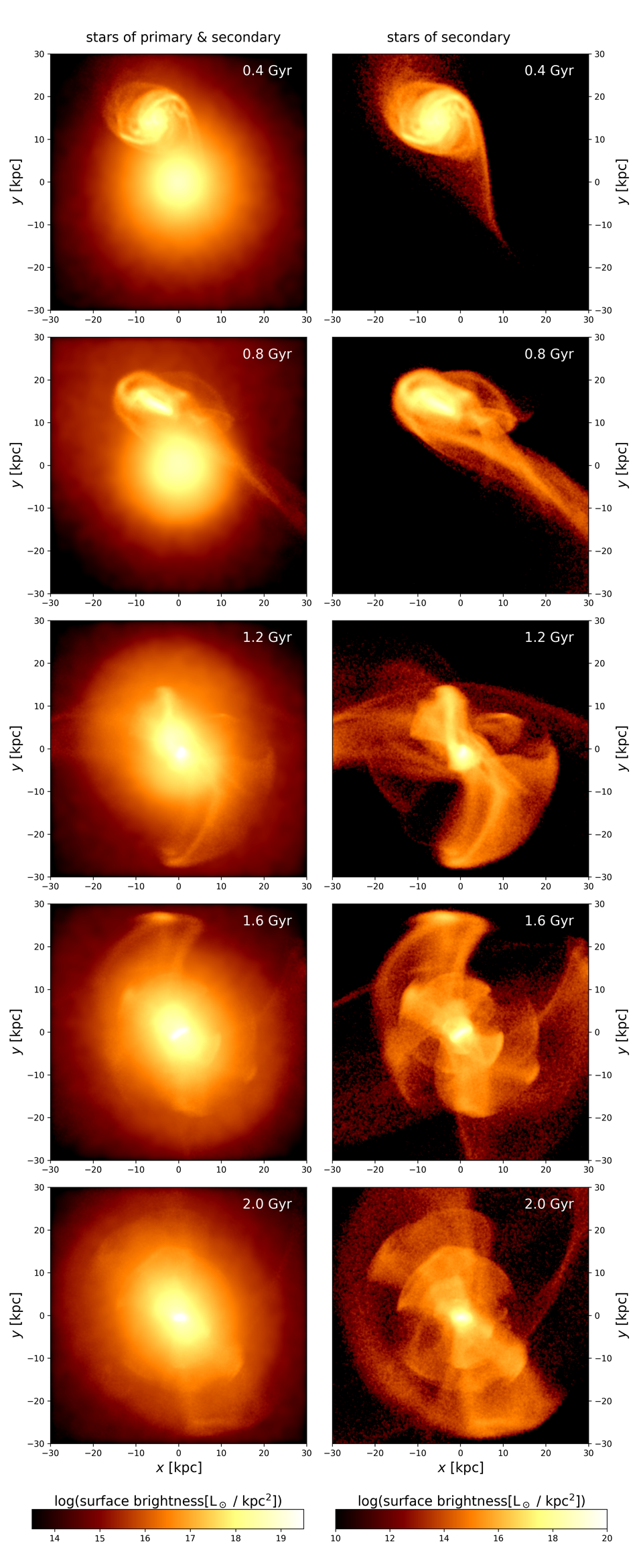}}
    \caption{Surface brightness maps at different stages of simulation.}
    \small Video available at: \url{https://doi.org/10.5281/zenodo.11247035}.
    \label{fig:snaps}
\end{figure}

Figure\,\ref{fig:snaps} shows surface brightness maps of several snapshots from the simulation. The projection plane corresponds to the merger plane, $xy$. 
The left column contains the light from the stellar components of both galaxies, in the right column only stellar particles that were originally in the disk of the secondary are included. 
The maps are computed in the same manner as the image described in Appx.\,\ref{apx:fits}.
Video of the whole simulation available at: \url{https://doi.org/10.5281/zenodo.11247035}.

\begin{figure*}
\resizebox{1.0\hsize}{!}{\includegraphics{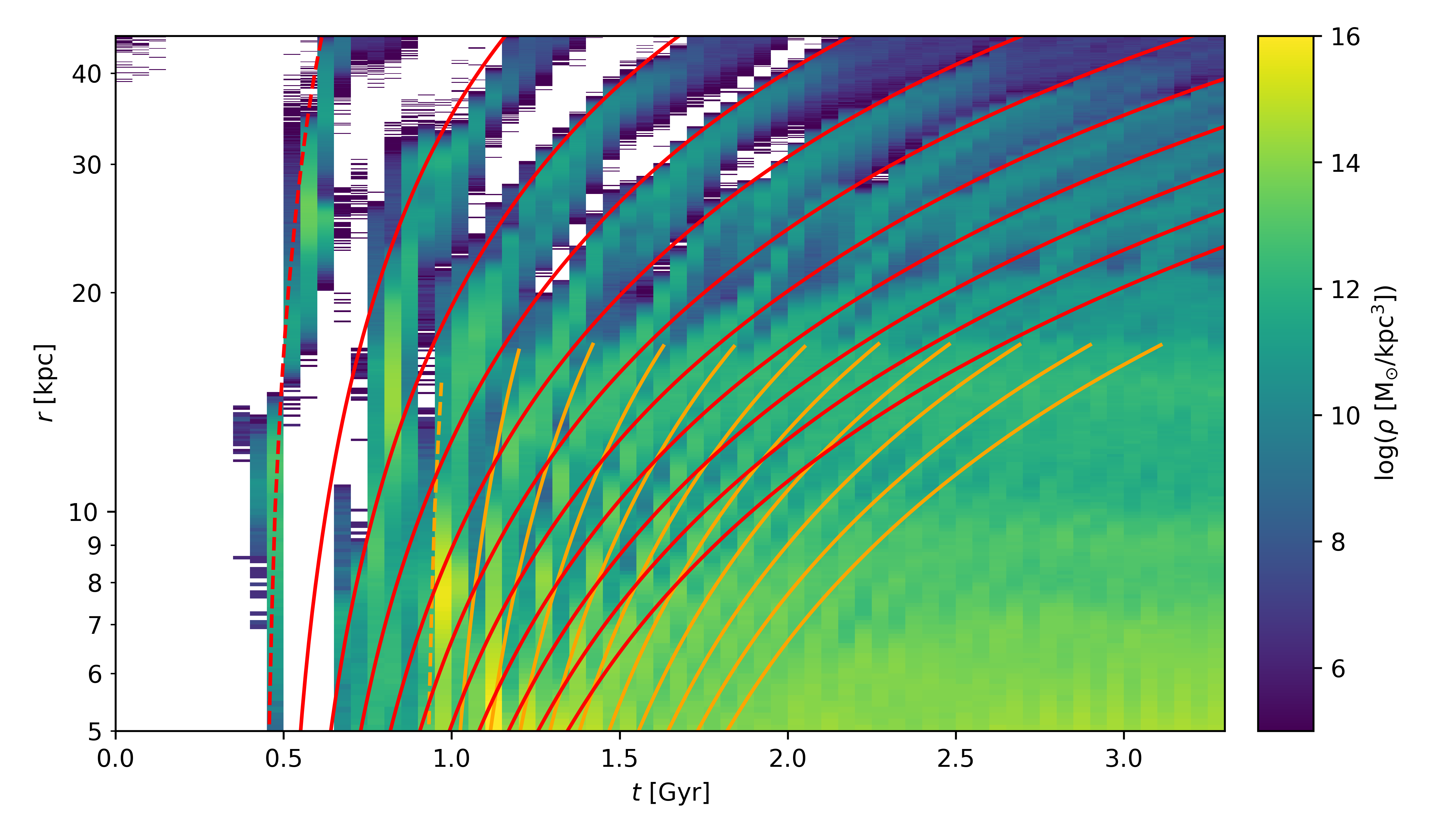}}
    \caption{Comparison of shell evolution in the simulation (the density histogram in color scale; shell edges correspond to the areas of the enhanced density) with our model (red and orange curves of the first and third shell generation, respectively; the so-called zeroth shells are marked with dashed curves). Time is measured from the beginning of the simulation, when the relative distance of the galaxies was 70\,kpc.}
    \label{fig:hist}
\end{figure*}

\subsection{Shell evolution} \label{sapx:evo}

Firstly, we wanted to compare the evolution of the shell radii in the simulation with the theoretical predictions of our model. The model assumes that shell-creating stars move on strictly radial orbits (see Sect.\,\ref{sec:evo}), while in the case of a nonradial merger with the disk of the secondary inclined to the orbital plane, one can expect stellar orbits of various shapes, which can cause the shell evolution in the simulation to deviate from the model prediction.
 
We calculated the analytic model of the shell evolution based on the gravitational potential 
measured at the snapshot 1.65\,Gyr after beginning of the simulation, neglecting the evolution of the potential. 
Each generation of the shells (see Sects.\,\ref{ssec:dis-1g} and \ref{ssec:dis-2g}) is assumed to originate at the moment of the respective pericentric passage of the galaxies. There were no adjustments made to match the model and the simulation. 

The simulated shells are tracked in the histogram of the time evolution of the density of the secondary stellar particles (i.e., particles that were in the disk of the secondary at the beginning of the simulation) in the spherical-shell bins. 
Shells consist of stars currently situated near their appocenters. 
To improve the visibility of the shell edges in the histogram, only particles with the galactocentric velocity (with respect to the center of the host, primary, galaxy) in the range $(-30:100)$\,km\,s$^{-1}$ are displayed. The positions of the shell edges correspond to the areas of the enhanced mass density.

Figure\,\ref{fig:hist} shows the comparison of the modeled shell evolution and the simulation. 
The outer shells in the simulation are well-represented, with the model of the first shell generation (red curves) released at the time of the first pericentric passage of the galaxies, 0.4\,Gyr after the beginning of the simulation. The deviation is in the range of several percent of the shell radii.
Shells at lower radii correspond better to the later shell generations -- in the figure, the third generation, released at 0.9\,Gyr, is represented by the orange curves.
The so-called zeroth shells (see Footnote\,\ref{fn:sh0}) are marked with dashed curves.

\subsection{Projection effects} \label{sapx:proj}

The histogram of the shell evolution in the simulation is averaged across the full solid angle. 
The effects of the nonradial merger, combined with the initial arrangement of the secondary stellar particles in the rotating disk and with the fact that the disk is inclined with respect to the merger plane, cause the 3D shape of the simulated shells to deviate from spherical caps. That means that the actual observed radius for the same shell can vary according to the projection plane and the position angle.
The variance depends on the details of the merger and it differs for different shells (e.g., shells with different serial numbers; see Sect.\,\ref{sec:evo}) as well as for the same shells in different stages of their evolution. 

A better quantification of the uncertainties coming from the projection effects would require extensive analysis of a wider set of simulations, which is beyond the scope of this work.
Here, we make an estimate using a snapshot of our simulation at 1.65\,Gyr. 
We measured shell radii in the surface density maps of $30\times30$\,kpc centered on the host galaxy, starting in the plane of the merger, $xy$, and continuing in planes rotated around $x$-axis by $10\degrees$ with respect to the previous one, ending after nine steps in the $xz$-plane.

The opening angle of different lines of sight, under which the substructure can be observed as a relatively sharp-edged shell, changes with time and with the shell serial number.
For the snapshot in hand, the outer (inner) shells are visible in the planes with the inclination up to 40--90$\degrees$ (20--30$\degrees$) with respect to the merger plane and their variations in radius are 5--19\,\% (2--14\,\%).

\section{Merger mass ratio estimates} \label{apx:fits}

\begin{figure}
\resizebox{1.0\hsize}{!}{\includegraphics{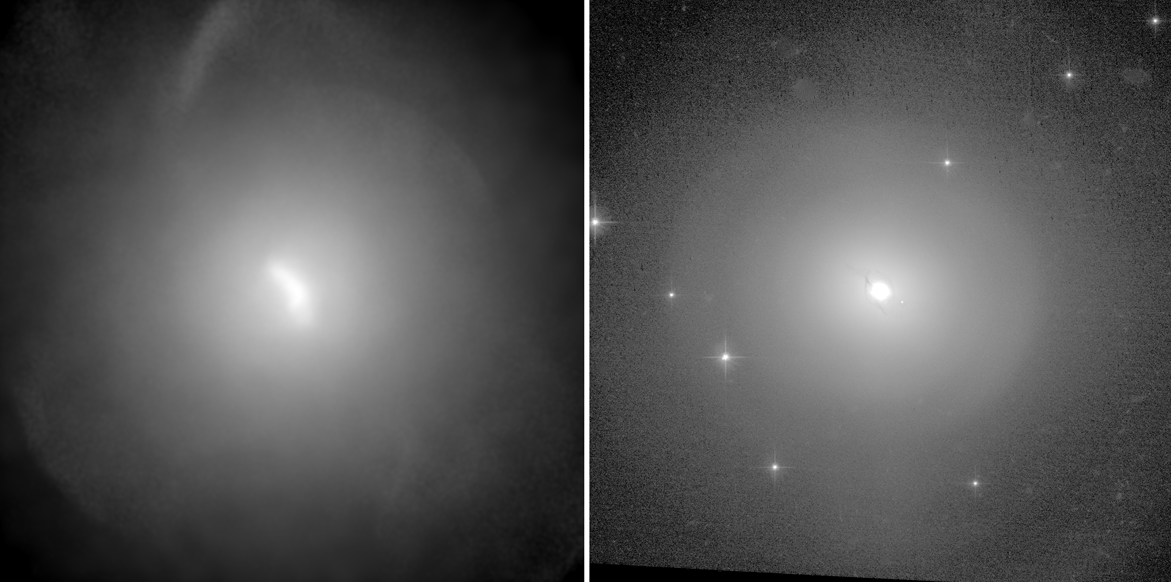}}
    \caption{Surface brightness map of the simulated galaxy (left) and the "cleaned image" of NGC\,4993, i.e., the HST image with faint stars and background galaxies removed (right). Those are the input data for the modeling of the host galaxy light. }
    \label{fig:fits}
\end{figure}

\begin{figure}
\resizebox{1.0\hsize}{!}{\includegraphics{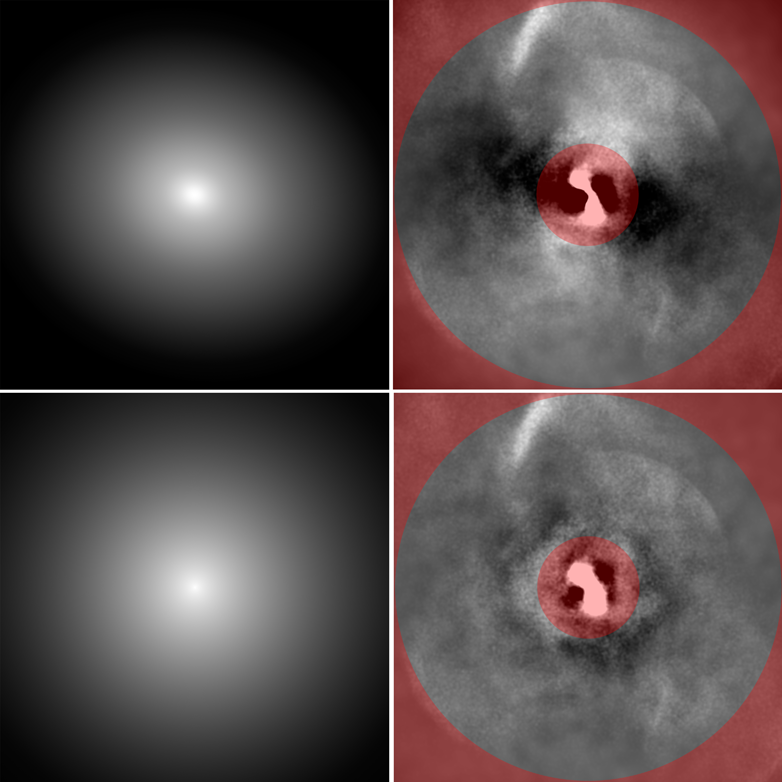}}
    \caption{Models of host galaxy light (left) and the residual images, i.e., the original image, left panel of Fig.\,\ref{fig:fits}, after the model subtraction (right) for the simulated galaxy. Top and bottom rows correspond to the IRAF Ellipse and GALFIT model, respectively, as described in Appx.\,\ref{apx:fits}. 
Areas excluded from the measurements of the signal in the residual images are shown in red. 
Model images have individual scales adjusted to the value range of the respective model. }
    \label{fig:mr_sim}
\end{figure}

\begin{figure}
\resizebox{1.0\hsize}{!}{\includegraphics{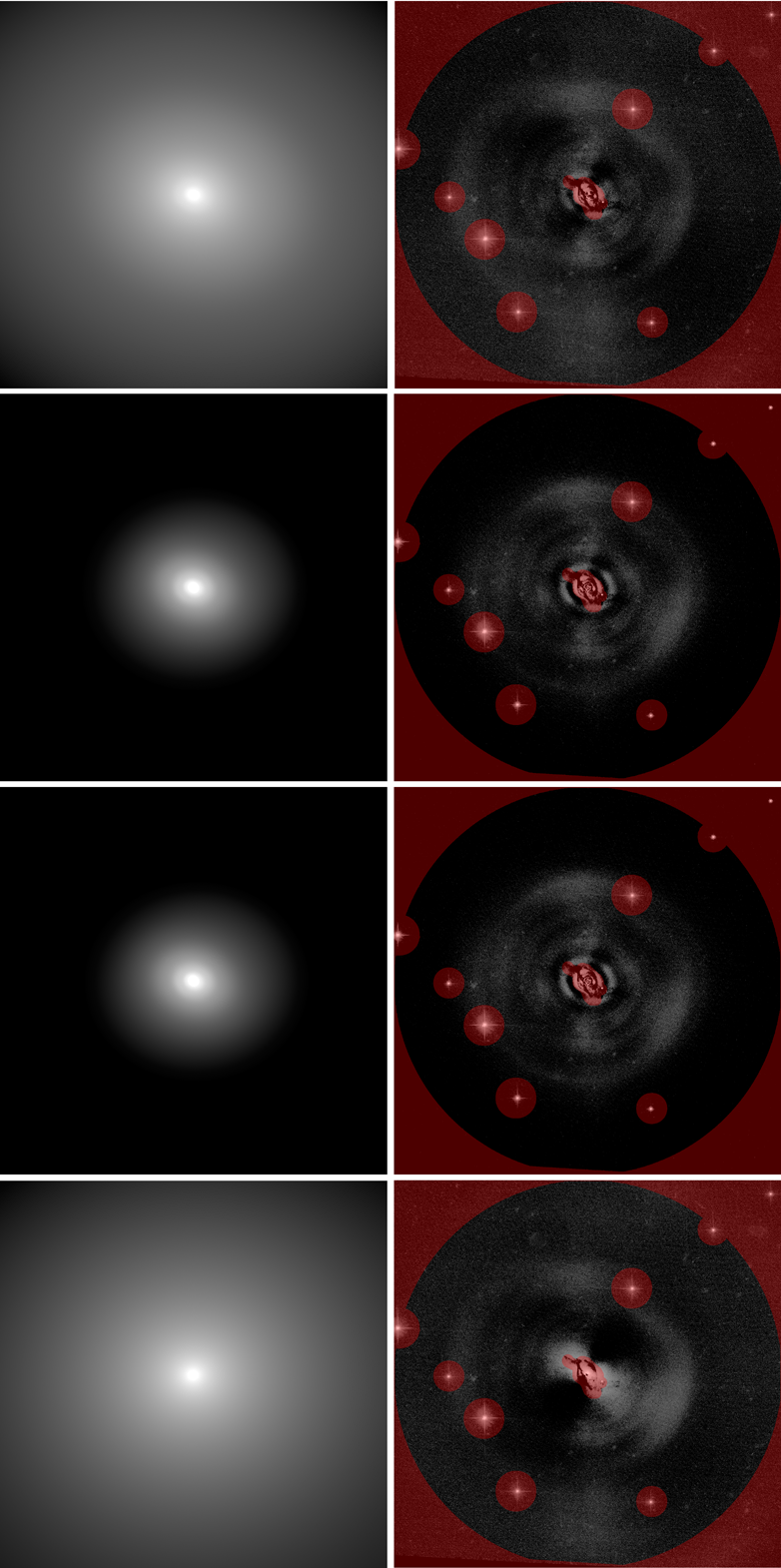}}
    \caption{Models of host galaxy light (left) and the residual images, i.e., the original image, right panel of Fig.\,\ref{fig:fits}, after the subtraction of the respective model (right) for the HST image of NGC\,4993. Rows correspond to models (1)\,--\,(4), respectively, as described in Appx.\,\ref{apx:fits}. 
Areas excluded from the measurements of the signal in the residual images are shown in red. 
Model images have individual scales adjusted to the value range of the respective model. }
    \label{fig:mr_hst}
\end{figure}

We attempt to estimate the merger mass ratio using the surface brightness map of the galaxy. In this approach, the main body of the host galaxy is approximated by a model, the model is subtracted from the original image, and the signal in the residual image is assumed to be proportional to the mass of the secondary galaxy.

Firstly, we want to estimate how much mass of the secondary is recovered in this way for the simulated galaxy.
We created a surface brightness map using the output of the simulation (Appx.\,\ref{apx:sim}) at 1.65\,Gyr, left panel of Fig.\,\ref{fig:fits}.
The field of view of the map is $40\times40$\,kpc (corresponding to $4096\times4096$ pixels) in the projection plane inclined by $-45\degrees$ with respect to the $zx$-plane around the $z$-axis, in other words, the projection plane is defined by the vectors (0, 0, 1) and (1, -1, 0), where $xy$ is the merger plane.
The mass of the particles is distributed over neighbouring pixels using a Gaussian kernel with the full width at half maximum equal to the gravitational softening length of the respective particles, which is 0.6 and 0.1\,kpc for the stellar particles of the primary and secondary, respectively.
For particles from the primary galaxy, we adopted a stellar mass-to-light ratio in the $I$-band $M_{*}/L=3$, similar to the values measured for the early-type galaxies of a similar absolute magnitude \citep{saur4}.
For the disk particles of the secondary, we opted for $M_{*}/L=1$ \citep{por04}.

We compared two different models of the brightness distribution of the host galaxy. The first one is created using IRAF Ellipse procedure with the fixed position angle and ellipticity. The other model is a single S{\'e}rsic profile fitted in GALFIT \citep{galfit}. We summed absolute values of the pixels between 5 and 19\,kpc from the center of the galaxy in the residual images. We avoid the central parts, because this region also needs to be excluded from the images of NGC\,4993 due to the significant dust absorption. 
Images of the models and residua are shown in  Fig.\,\ref{fig:mr_sim}.

The ratio of the signal in the residual and original image in this area is 0.11 and 0.05 for the IRAF and GALFIT models of the host galaxy, respectively. 
In our approach, this represents the brightness fraction (in the given area) coming from the stars of the secondary galaxy, and the fraction is derived in a way applicable to the observed photometric data. 
The actual stellar-mass fraction of the secondary particles in the area is 0.07. With the adopted mass-to-light ratios for the stellar particles, this translates to the brightness fraction of 0.19. 
Note that the overall stellar-mass fraction of the secondary was set to 0.09 at the beginning of the merger simulation.

We applied a similar procedure to the HST image of NGC\,4993. Firstly, faint stars and background galaxies were removed to obtain a "cleaned image", shown in the right panel of Fig.\,\ref{fig:fits}, then bright stars and the central dusty region were masked out from the image.
We constructed four different models of the host galaxy:
(1) IRAF Ellipse with the fixed position angle and ellipticity, the step size for the ellipse fitting of 0.4 pixels, and three sigma clippings applied three times;
(2) IRAF Ellipse with the position angle and ellipticity parameters free and the same step size and sigma clipping as the first model;
(3) IRAF Ellipse with the same parameters as the second model but five sigma clippings applied ten times;
(4) GALFIT with a single S{\'e}rsic profile combined with a linear sky gradient.
Images of the models and residua are shown in  Fig.\,\ref{fig:mr_hst}.
We summed absolute values of the pixels up to $45.2\arcsec$ from the center of the galaxy in the residual images, excluding the masked areas.
The brightness ratio of the sum to the signal in the same area in the "cleaned image" is 0.13, 0.27, 0.28, and 0.14 for the models (1)\,--\,(4), respectively.

The methods described above recover the brightness ratios for NGC\,4993 to be roughly twice as large as in the case of the simulated galaxy.
Given the different stellar mass-to-light ratios of the primary and secondary, this leads us to a rough estimate, that the galaxy accreted on NGC\,4993 was 2-3 times more massive in the stellar component than the secondary galaxy in the simulation. Note that the initial stellar-mass ratio in the simulation was 1:10.

It is precarious to draw conclusions from the analysis of one snapshot from one simulation.
We checked that the stellar-mass fraction of the secondary in the concerned area stays between 0.7 and 0.8 for a 1\,Gyr period around the concerned output of the simulation, although it may result in different values in the residual images of different outputs.
To acquire more robust estimates would require extensive analysis of a wider set of simulations, which is beyond the scope of this paper, and it is going to be the subject of the future work.
It is important to note that the merger mass ratio is one of the parameters used in the estimate of the probability that the progenitor of GW170817 originated in the accreted galaxy, and the mass ratio does not directly influence our estimate of the time since the galactic merger.


\end{document}